\documentclass[12pt]{iopart}
\begin{document}
\title{Factorization solution of Cari\~nena's quantum nonlinear oscillator.}
\author{Jonathan M Fellows and Robert A Smith}
\address{School of Physics and Astronomy, University of Birmingham, Edgbaston, Birmingham B15 2TT, United Kingdom}
\eads{\mailto{fellowsjm@theory.bham.ac.uk}, \mailto{ras@th.ph.bham.ac.uk}}
\begin{abstract}
In a recent paper \cite{CPRS08} Cari\~nena et al analyzed a non-polynomial one-dimensional quantum potential representing an oscillator which they argued was intermediate between the harmonic and isotonic oscillators. In particular they proved that it is Schr\"odinger soluble, and explicitly obtained the wavefunctions and energies of the bound states. In this paper we show that these results can be obtained much more simply by noting that this potential is a supersymmetric partner potential of the harmonic oscillator. We then use this observation to generate an infinite set of potentials which can be exactly solved in a similar manner.
\end{abstract}

\section{Introduction}

In a recent paper in this journal, Cari\~nena et al \cite{CPRS08} investigated the solutions to the eigenvalue problem for the Schr\"odinger equation \cite{Factor of 2}
\begin{equation*}
-{d^2\psi\over dx^2}+\left[x^2+8{2x^2-1\over (2x^2+1)^2}\right]\psi=E\psi.
\end{equation*}
Using some rather involved mathematics they were able to show that the eigenfunctions are given by
\begin{equation*}
\psi_n(x)={P_n(x)\over (2x^2+1)}e^{-x^2/2},
\qquad\hbox{where}\qquad n=0,3,4,5\dots,
\end{equation*}
and the polynomial factors $P_n(x)$ are related to the Hermite polynomials by
\begin{equation*}
P_n(x)=
\cases{
1&n=0\cr
H_n(x)+4nH_{n-2}(x)+4n(n-3)H_{n-4}(x)&n=3,4,5\dots\cr}
\end{equation*}
The corresponding eigenvalues are given by
\begin{equation*}
E_n=-3+2n\qquad\hbox{where}\qquad n=0,3,4,5\dots
\end{equation*}
In this paper we show that Cari\~nena's potential is a supersymmetric partner potential of the harmonic oscillator. This allows us to rederive all the above results in a much simpler fashion. In addition, by considering the question of which other potentials are supersymmetric partners of the harmonic oscillator, we construct an infinite set of exactly soluble potentials, along with their eigenfunctions and eigenvalues.

The rest of the paper is organized as follows. In section 2 we give a brief summary of the ideas of the factorization approach to the Schr\"odinger equation and supersymmetric partner potentials. In section 3 we prove that Cari\~nena's potential is a partner potential of the harmonic oscillator, and use this to rederive the formulae for its eigenfunctions and eigenvalues. In section 4 we show how to find other partner potentials of the harmonic oscillator, and derive an infinite set of exactly soluble potentials.

\section{The factorization approach}

In this section we provide a self-contained introduction to the factorization approach, and the idea of supersymmetric partner potentials. More details can be found in references
\cite{Cooper 2001} and \cite{IH51}.

We start with the one-dimensional single-particle Schr\"odinger equation,
\begin{equation*}
H_1\psi(x)=\left[-{d^2\over dx^2}+V_1(x)\right]\psi(x)=E\psi(x).
\end{equation*}
The idea is to factorize the Hamiltonian operator, $H_1$, which is a second-order differential operator, into a product of two first-order differential operators,
\begin{equation*}
H_1=A^{\dagger}A,
\end{equation*}
where
\begin{equation*}
A={d\over dx}+W(x),\qquad
A^{\dagger}=-{d\over dx}+W(x).
\end{equation*}
Upon direct substitution we see that
\begin{equation*}
V_1(x)=W^2(x)-W'(x).
\end{equation*}
We now define the operator $H_2=AA^{\dagger}$ by reversing the order of $A$ and $A^{\dagger}$. Simple algebra shows that $H_2$ is a Hamiltonian corresponding to a new potential $V_2(x)$,
\begin{equation*}
H_2=AA^{\dagger}=-{d^2\over dx^2}+V_2(x),\qquad
V_2(x)=W(x)^2+W'(x).
\end{equation*}
The potentials $V_1(x)$ and $V_2(x)$ are known as supersymmetric partner potentials.

The key result we will be using in later sections is that the eigenvalues and eigenfunctions of $H_1$ and $H_2$ are related. Suppose that $\psi^{(1)}_n$ is an eigenfunction of $H_1$ with energy eigenvalue $E^{(1)}_n$. Then $A\psi^{(1)}_n$ is an eigenfunction of $H_2$ with energy eigenvalue $E^{(1)}_n$ since
\begin{equation*}
H_2\left[A\psi^{(1)}_n\right]=
AA^{\dagger}A\psi^{(1)}_n=
A\left[H_1\psi^{(1)}_n\right]=
A\left[E^{(1)}_n\psi^{(1)}_n\right]=
E^{(1)}_n\left[A\psi^{(1)}_n\right].
\end{equation*}
The only way that $A\psi^{(1)}_n$ could fail to be an eigenfunction of $H_2$ is if it equals zero, in which case
$\displaystyle W(x)=-{d\over dx}\ln\psi^{(1)}_n(x)$, which could be true for at most one value of $n$. Similarly suppose that $\psi^{(2)}_n$ is an eigenfunction of $H_2$ with eigenvalue $E^{(2)}_n$. Then $A^{\dagger}\psi^{(2)}_n$ is an eigenfunction of $H_1$ with energy eigenvalue $E^{(2)}_n$ since
\begin{equation*}
H_1\left[A^{\dagger}\psi^{(2)}_n\right]=
A^{\dagger}AA^{\dagger}\psi^{(2)}_n=
A^{\dagger}\left[H_2\psi^{(2)}_n\right]=
A^{\dagger}\left[E^{(2)}_n\psi^{(2)}_n\right]=
E^{(2)}_n\left[A^{\dagger}\psi^{(2)}_n\right].
\end{equation*}
The only way that $A^{\dagger}\psi^{(2)}_n$ could fail to be an eigenfunction of $H_1$ is if it equals zero, in which case
$\displaystyle W(x)={d\over dx}\ln\psi^{(2)}_n(x)$, which again could be true for at most one value of $n$.

It follows that, with the possible exception of one level, $H_1$ and $H_2$ have the same energy spectra. The eigenfunctions of $H_2$ can be found from those of $H_1$ by applying the operator $A$; the eigenfunctions of $H_1$ can be obtained from those of $H_2$ by applying the operator $A^{\dagger}$. If $H_1$ has one more energy level than $H_2$, this energy eigenfunction will be the solution of $A\psi^{(1)}_0=0$, and have energy eigenvalue zero; if $H_2$ has one more energy level than $H_1$, this energy eigenfunction will be a solution of $A^{\dagger}\psi^{(2)}_0=0$, and have energy eigenvalue zero. The upshot of all this is that, if we know how to exactly solve one of $H_1$ or $H_2$, we can immediately derive the exact solution of the other.

Finally we note that we can easily normalize the newly obtained eigenfunctions. If $\psi^{(1)}_n$ is a normalized eigenfunction of $H_1$ with energy $E^{(1)}_n$, the corresponding normalized eigenfunction of $H_2$ is
\begin{equation*}
\psi^{(2)}_n=C_n A \psi^{(1)}_n,
\end{equation*}
where $C_n$ is a constant to be determined. If we take the scalar product of this eigenfunction with itself we get
\begin{equation*}
1=(\psi^{(2)}_n,\psi^{(2)}_n)=
|C_n|^2 (\psi^{(1)}_n,A^{\dagger}A\psi^{(1)}_n)=
|C_n|^2 E^{(1)}_n.
\end{equation*}
This leads to the result
\begin{equation*}
\psi^{(2)}_n=\left[E^{(1)}_n\right]^{-1/2}A\psi^{(1)}_n
=\left[E^{(1)}_n\right]^{-1/2}\left[{d\over dx}+W(x)\right]
\psi^{(1)}_n,
\end{equation*}
and we similarly find that
\begin{equation*}
\psi^{(1)}_n=\left[E^{(2)}_n\right]^{-1/2}A^{\dagger}\psi^{(2)}_n
=\left[E^{(2)}_n\right]^{-1/2}\left[-{d\over dx}+W(x)\right]
\psi^{(2)}_n.
\end{equation*}

\section{Solution of the Cari\~nena potential}

We can solve the Cari\~nena potential by noticing that it is a partner potential of the harmonic oscillator. To see this we simply take
\begin{equation*}
W(x)=x+{4x\over (2x^2+1)},
\end{equation*}
from which trivial calculus gives us
\begin{eqnarray*}
V_1(x)&=W(x)^2-W'(x)= x^2 + 8{(2x^2-1)\over (2x+1)^2} + 3\\
V_2(x)&=W(x)^2+W'(x)= x^2 + 5.
\end{eqnarray*}
If follows that
\begin{eqnarray*}
H_1&=A^{\dagger}A=H_C+3\quad\rightarrow\quad
E_k^{(1)}=E_k^C+3\\
H_2&=AA^{\dagger}=H_H+5\quad\rightarrow\quad
E_k^{(2)}=E_k^H+5,
\end{eqnarray*}
where
\begin{equation*}
H_C=-{d^2\over dx^2}+x^2+8{(2x^2-1)\over (2x^2+1)^2}
\end{equation*}
is the Cari\~nena Hamiltonian, and
\begin{equation*}
H_H=-{d^2\over dx^2}+x^2
\end{equation*}
is the harmonic oscillator Hamiltonian.

Now the eigenvalues and unnormalized eigenfunctions of the harmonic oscillator \cite{Flugge 1994} are given by
\begin{equation*}
E^H_k=2k+1\qquad\hbox{and}\qquad
\phi^H_k(x)=H_k(x)e^{-x^2/2},
\end{equation*}
where $k=0,1,2\dots$ and the Hermite polynomials $H_k(x)$ are given by the Rodrigues formula
\begin{equation*}
H_k(x)=(-1)^k e^{x^2} {d^k\over dx^k} \left[e^{-x^2}\right].
\end{equation*}
If follows trivially that the eigenvalues and unnormalized eigenfunctions of $H_2$ are
\begin{equation*}
E^{(2)}_k=2k+6\qquad\hbox{and}\qquad
\psi^{(2)}_k(x)=H_k(x)e^{-x^2/2}.
\end{equation*}
The corresponding eigenvalues of $H_1$ are therefore $E^{(1)}_k=2k+6$, with unnormalized eigenfunctions
\begin{equation*}
\psi^{(1)}_k(x)=A^{\dagger}\psi^{(2)}_k(x)=
\left[-{d\over dx}+x+{4x\over (2x+1)}\right]
H_k(x)e^{-x^2/2}.
\end{equation*}
We can simplify this expression by repeatedly using the Hermite polynomial identities
\begin{eqnarray*}
H_k'(x)&=2k H_{k-1}(x)\\
2xH_k(x)&=H_{k+1}(x)+2kH_{k-1}(x),
\end{eqnarray*}
which can be derived directly from the Rodrigues formula, to obtain
\begin{equation*}
\psi^{(1)}_k(x)={1\over 2}\left[H_{k+3}(x)+4(k+3)H_{k+1}(x)
+4k(k+3)H_{k-1}(x)\right]{e^{-x^2/2}\over (2x^2+1)}
\end{equation*}
The normalization factor for the harmonic oscillator eigenfunctions is
\begin{equation*}
N^{(2)}_k=\left[{1\over 2^k k!\sqrt{\pi}}\right]^{1/2},
\end{equation*}
so the corresponding factor for the Cari\~nena eigenfunctions is
\begin{equation*}
N^{(1)}_k=\left[E^{(2)}_k\right]^{-1/2}N^{(2)}_k
=\left[{1\over 2^{k+1} k! (k+3)\sqrt{\pi}}\right]^{1/2}.
\end{equation*}
We have therefore exactly solved the Cari\~nena Hamiltonian, and the eigenvalues and normalized eigenfunctions are given by
\begin{eqnarray*}
E^C_k&=2k+3\\
\phi^C_k(x)&=\left[(k+1)(k+2)\over 2^{k+3}(k+3)!\sqrt{\pi}\right]^{1/2}
{\left[H_{k+3}(x)+4(k+3)H_{k+1}(x)+4k(k+3)H_{k-1}(x)\right]
\over (2x^2+1)}e^{-x^2/2}
\end{eqnarray*}
where $k=0,1,2\dots$ To compare with Cari\~nena et al's results we set $n=k+3$ so that
\begin{eqnarray*}
E^C_n&=2n-3\\
\phi^C_n(x)&=\left[(n-1)(n-2)\over 2^n n! \sqrt{\pi}\right]^{1/2}
{\left[H_{n}(x)+4nH_{n-2}(x)+4n(n-3)H_{n-4}(x)\right]
\over (2x^2+1)}e^{-x^2/2},
\end{eqnarray*}
where $n=3,4,5\dots$ Note that there is one final state possible, which is the solution of $A\phi_0=0$, and this will have energy $-3$ if it exists. Solving this equation gives
\begin{equation*}
\psi^C_0(x)=C{e^{-x^2/2}\over (2x^2+1)},
\end{equation*}
which is of exactly the same form as the previous $\phi^C_n(x)$ with $n=0$. Even the form of the normalisation constant suggested by placing $n=0$ in the previous equation turns out to be correct, as we show in Appendix A. It follows that the final solution of the Cari\~nena Hamiltonian is
\begin{eqnarray*}
E^C_n&=2n-3\\
\phi^C_n(x)&=\left[(n-1)(n-2)\over 2^n n! \sqrt{\pi}\right]^{1/2}
{\left[H_{n}(x)+4nH_{n-2}(x)+4n(n-3)H_{n-4}(x)\right]
\over (2x^2+1)}e^{-x^2/2},
\end{eqnarray*}
where $n=0,3,4,5\dots$

We have therefore reproduced the results of Cari\~nena et al in a very direct and systematic manner by using the fact that the Cari\~nena potential is a partner potential of the harmonic oscillator potential.

\section{Partner potentials of the harmonic oscillator}

The solution of the Cari\~nena Hamiltonian in section 3 was motivated by the chance observation of one of the authors (JMF) that it is a partner of the harmonic oscillator. Let us now turn this around, and ask which Hamiltonians, $H_1$, are partners to the harmonic oscillator, $V_2(x)=x^2$, and can thus be solved by the method of section 3. The $W(x)$ needed in the factorisation method would then have to satisfy
\begin{equation*}
{dW\over dx}+W(x)^2=V_2(x)-\lambda,
\end{equation*}
where we have included an irrelevant constant $\lambda$. This is an example of a Riccati differential equation \cite{Ince 1956}, first analysed in 1724. The method of solution is to substitute
\begin{equation*}
W(x)={d\over dx}\ln{\phi(x)}={1\over\phi(x)}{d\phi\over dx},
\end{equation*}
which leads to the equation
\begin{equation*}
-{d^2\phi\over dx^2}+V_2(x)\phi(x)=\lambda\phi(x).
\end{equation*}
We see that $\phi(x)$ is a solution of the original Schr\"odinger equation, in our case the harmonic oscillator. We seem to have gone round in a complete circle, which is not surprising since the central idea of the partner potential method is a mapping between second order linear and first order non-linear differential equations.

We also seem to run into a problem in that every solution $\phi_k(x)$ of the original Schr\"odinger equation, other than the ground state, will have at least one zero $x_0$. The $W(x)$ generated from this will then have a $(x-x_0)^{-1}$ singularity, and then $V_1(x)$ will have a $(x-x_0)^{-2}$ singularity. This approach would therefore seem to only generate soluble potentials which have a singularity at finite $x$, which might be considered an unphysical feature.

This problem can be overcome in the special case of the harmonic oscillator, for which the Schr\"odinger equation is
\begin{equation*}
-{d^2\phi\over dx^2}+x^2\phi=\lambda\phi.
\end{equation*}
If we set $x=iy$, this becomes
\begin{equation*}
-{d^2\phi\over dy^2}+y^2\phi=-\lambda\phi,
\end{equation*}
which is the original equation with the irrelevant change in constant $\lambda\rightarrow -\lambda$. In other words, if we set $x\rightarrow ix$ in the harmonic oscillator eigenfunctions, we get perfectly good $\phi(x)$ which can then generate $W(x)$ and finally $H_1$ and $H_2$. These $\phi(x)$ would not be good eigenfunctions as they are not normalizable -- they behave like $e^{x^2/2}$ at large $x$ -- but this is not relevant here. The solutions $\phi(x)$ are thus
\begin{equation*}
\phi_p(x)={\cal H}_p(x) e^{x^2/2},
\end{equation*}
where the pseudo-Hermite polynomials are given by
\begin{equation*}
{\cal H}_p(x)=(-i)^p H_p(ix)
=e^{-x^2}{d^p\over dx^p}\left[e^{x^2}\right].
\end{equation*}
They are basically the Hermite polynomials, where the signs of all coefficients are made positive. The even solutions $\phi_{2m}(x)$ have no real zeros; the odd solutions $\phi_{2m+1}(x)$ have their only real zero at $x=0$ since they are odd. This leads to a $1/x^2$ singularity at $x=0$, which can be regarded as a centrifugal barrier, as in the case of the isotonic oscillator.

Let us examine the first few partner potentials generated in this manner. For $k=0$,
\begin{eqnarray*}
\phi_0(x)&=e^{x^2/2}\\
W(x)&={d\over dx}\ln{\phi_0(x)}=x\\
V_1(x)&=x^2-1\\
V_2(x)&=x^2+1,
\end{eqnarray*}
and this generates the standard ladder operator solution of the harmonic oscillator, since this shows that the harmonic oscillator is a partner to itself shifted by 2 units of energy.
For $k=1$,
\begin{eqnarray*}
\phi_1(x)&=x e^{x^2/2}\\
W(x)&={d\over dx}\ln{\phi_1(x)}=x+{1\over x}\\
V_1(x)&=x^2+{2\over x^2}+1\\
V_2(x)&=x^2+3,
\end{eqnarray*}
and the partner potential is an example of the isotonic oscillator. The eigenfunctions of the isotonic oscillator generated by this method,
\begin{equation*}
\psi^{(1)}_m(x)=\left[-{d\over dx}+x+{1\over x}\right]H_m(x)e^{-x^2/2},
\end{equation*}
are only non-singular for odd $m$, and hence the energy spectrum is $E_k=2k+2$ where $k=1,3,5\dots$, so that the level spacing is double that of the harmonic oscillator.
\par\noindent
For $k=2$,
\begin{eqnarray*}
\phi_2(x)&=(4x^2+2) e^{x^2/2}\\
W(x)&={d\over dx}\ln{\phi_1(x)}=x+{4x\over (2x^2+1)}\\
V_1(x)&=x^2+8{(2x^2-1)\over (2x^2+1)^2}+3\\
V_2(x)&=x^2+5,
\end{eqnarray*}
and this is the Cari\~nena potential we have just solved. This is the first new soluble potential generated by this method, since the solutions of the harmonic and isotonic oscillator are well-known.
\par\noindent
For $k=3$,
\begin{eqnarray*}
\phi_3(x)&=(8x^3+12x) e^{x^2/2}\\
W(x)&={d\over dx}\ln{\phi_3(x)}=x+{1\over x}+{4x\over (2x^2+3)}\\
V_1(x)&=x^2+{2\over x^2}+8{(2x^2-3)\over (2x^2+3)^2}+5\\
V_2(x)&=x^2+7,
\end{eqnarray*}
which is a variant of the Cari\~nena potential which includes a centrifugal barrier term.
\par\noindent
For $k=4$,
\begin{eqnarray*}
\phi_4(x)&=(16x^4+24x^2+12) e^{x^2/2}\\
W(x)&={d\over dx}\ln{\phi_3(x)}=x+{(16x^3+24x)\over (4x^4+12x^2+3)}\\
V_1(x)&=x^2+16{(8x^6+12x^4+18x^2-9)\over (4x^4+12x^2+3)^2}+7\\
V_2(x)&=x^2+9,
\end{eqnarray*}
which is the harmonic potential plus a rational potential which is regular at the origin, and falls off at infinity.

The common feature of these potentials is that they consist of a harmonic term plus an additional rational function which falls off at infinity like
a constant times $1/x^2$. For even $p$ this additional term is symmetric and finite at the origin, and leads to a potential which looks like a
harmonic well with an attractive dimple. For odd $p$ the additional term is similar to that for even $p-1$, plus a centrifugal barrier term. The latter means that we consider these potentials in the interval $(0,\infty)$, as in the case of the isotonic oscillator.
 
\section{Exact solution of Generalized Cari\~nena potentials}

In this section we will write down the exact closed form solutions of the Generalized Cari\~nena potentials, $V_C^p(x)$, which we define to be the partner potentials of the harmonic oscillator generated by the functions $\phi_p(x)$ introduced in the last section. The original Cari\~nena potential corresponds to the case $p=2$.

We first deduce an expression for $V_C^p(x)$. Since
\begin{eqnarray*}
W(x)&={d\over dx}\ln{\phi_p(x)}={\phi'_p(x)\over\phi_p(x)}
=x+{{\cal H}'_p(x)\over{\cal H}_p(x)}\cr
W'(x)&={\phi''_p(x)\over\phi_p(x)}-
\left[{\phi'_p(x)\over\phi_p(x)}\right]^2
=1+{{\cal H}''_p(x)\over{\cal H}_p(x)}
-\left[{{\cal H}'_p(x)\over{\cal H}_p(x)}\right]^2,
\end{eqnarray*}
it follows that
\begin{eqnarray*}
\fl
V_2(x)&=W^2(x)+W'(x)={\phi''_p(x)\over\phi_p(x)}=x^2+2p+1\cr
\fl
V_1(x)&=W^2(x)-W'(x)=V_2(x)-2W'(x)=
x^2+2{{\cal H}'_p(x)^2-
{\cal H}_p(x){\cal H}''_p(x)\over {\cal H}_p(x)^2}+2p-1.
\end{eqnarray*}
We therefore deduce the formula for the Generalized Cari\~nena potential,
\begin{equation*}
V_C^p(x)=x^2+2{{\cal H}'_p(x)^2-
{\cal H}_p(x){\cal H}''_p(x)\over {\cal H}_p(x)^2}.
\end{equation*}
Since ${\cal H}_p(x)$ is a polynomial of order $p$, we see that the second term in $V_C^p(x)$ is a rational function with numerator a polynomial of degree $2p-2$, and denominator a polynomial of degree $2p$. As $x\rightarrow\infty$,
\begin{equation*}
V_C^p(x)\sim x^2+{2p\over x^2}.
\end{equation*}

Starting from the original harmonic oscillator eigenvalues,
$E^H_k=2k+1$, then $E^{(1)}_k=E^{(2)}_k=2(k+p+1)$ and
$E^C_k=2k+3$. The unnormalized eigenfunctions are
\begin{eqnarray*}
\fl
\psi^{(1)}_k(x)&=A^{\dagger}\psi^{(2)}(x)=
\left[-{d\over dx}+x+{{\cal H}'_p(x)
\over{\cal H}_p(x)}\right]H_k(x)e^{-x^2/2}\cr
\fl
&=\left[2xH_k(x)-H'_k(x)+{{\cal H}'_p(x)
\over{\cal H}_p(x)}H_k(x)\right]e^{-x^2/2}\cr
\fl
&=\left[H_{k+1}(x)+{{\cal H}'_p(x)
\over{\cal H}_p(x)}H_k(x)\right]e^{-x^2/2}\cr
\fl
&=\left[{\cal H}_p(x)H_{k+1}(x)+
{\cal H}'_p(x)H_k(x)\right] {e^{-x^2/2}\over{\cal H}_p(x)}\cr
\fl
&=\left[\sum_{i=0}^p 2^i {p!\over i!(p-i)!}\{k+p+1\}
{(k+p-i)!\over (k+p+1-2i)!}H_{k+p+1-2i}(x)\right]
{e^{-x^2/2}\over{\cal H}_p(x)}.
\end{eqnarray*}
The normalization factor is given by
\begin{equation*}
N^{(1)}_k=\left[E^{(2)}_k\right]^{-1/2}N^{(2)}_k
=\left[{1\over 2^{k+1}k!(k+p+1)\sqrt{\pi}}\right]^{1/2}
\end{equation*}
If we define $n=k+p+1$ we see that the eigenfunctions are given by
\begin{equation*}
\phi^C_n(x)=N_n{P_n(x)\over{\cal H}_p(x)}e^{-x^2/2},
\qquad\hbox{where}\qquad
n=p+1,p+2,p+3\dots,
\end{equation*}
the polynomial factors $P_n(x)$ are related to the Hermite polynomials by
\begin{equation*}
P_n(x)=\sum_{i=0}^p {p!\over i!(p-i)!} n
{(n-i-1)!\over (n-2i)!}H_{n-2i}(x),
\end{equation*}
and the normalization constant $N_n$ is
\begin{equation*}
N_n=\left[{(n-1)(n-2)\dots(n-p)\over 2^{n-p}\,n!\,\sqrt{\pi}}\right]^{1/2}
\end{equation*}
The corresponding eigenvalues are given by
\begin{equation*}
E^C_n=-2p+1+2n
\qquad\hbox{where}\qquad
n=p+1,p+2,p+3\dots
\end{equation*}
There is one final state possible, which is given by $A\psi^{(1)}_0=0$, and has energy $E^{(1)}_0=0$, so that $E^C_0=-2p+1$. This state is then given by
\begin{equation*}
\phi_0^C(x)={1\over\phi_p(x)}=N_0{e^{-x^2/2}\over{\cal H}_p(x)}
\end{equation*}
When $p=2m$ is even, the normalization constant can be deduced from Appendix A to be
\begin{equation*}
N_0=\left[(2m)!2^m\over \sqrt{\pi}\right]^{1/2},
\end{equation*}
which is exactly what we would obtain if we naively set $n=0$ in our equation for the normalization constant $N_n$. It follows that for even $p$, all the previous formulas are still correct, but now we have $n=0,p+1,p+2,p+3,\dots$, so that we have an equidistant level structure but with $p$ levels missing.

In the case where $p$ is odd, every other level starting with the ground state will have a $1/x$ singularity at $x=0$ and hence will not be square normalizable, and should be eliminated. This gives an equidistant spectrum with twice the spacing of the harmonic oscillator levels, exactly as in the case of the isotonic oscillator.

\section{Summary and Conclusions}

We have obtained the exact solution of Cari\~nena's quantum non-linear oscillator in an economical fashion using the methods of supersymmetric quantum mechanics to show that it is a partner of the harmonic oscillator. We have then used this approach to define a countably infinite set of generalized Cari\~nena potentials $V^C_p(x)$, where $p$ is a positive integer, which we then exactly solve in a similar fashion. The case where $p$ is even is the most interesting, since all the generated eigenfunctions are normalizable. When $p$ is odd, half of the generated eigenfunctions must be removed as they are not normalizable.

The key technical observation is that the harmonic oscillator eigenvalue equation is unchanged under the transformation $x\rightarrow ix$, so the harmonic oscillator eigenfunctions with $x$ replaced by $ix$ can be used to generate the ladder operators $A$ and $A^{\dagger}$ needed to move between supersymmetric partners. More generally one can use any solution of a Schr\"odinger equation to generate the ladder operators; it does not have to be a good eigenfunction as normalization is not required.

We finally note that the isotonic oscillator eigenvalue equation is also unchanged under the transformation $x\rightarrow ix$, and this will be the subject of future investigations.

\ack

We would like to thank M W Long and T W Silk for useful comments.
RAS is supported by the UK EPSRC under Grant EP/D031109.

\appendix
\section{Normalization of ground state eigenfunctions}

In this appendix we evaluate the normalization integral needed for the ground state eigenfunctions
\begin{equation}
\label{IntIdent}
I_{2m}=\int_{-\infty}^{\infty} {e^{-x^2}\over{\cal H}_{2m}(x)^2}\,dx
={\sqrt{\pi}\over 2^{2m}(2m)!}
\end{equation}
To evaluate this integral define $J_p(x)$ as the indefinite integral
\begin{equation*}
J_p(x)=\int {e^{-x^2}\over{\cal H}_p(x)^2}\,dx
=\int G_p(x)^2 e^{-x^2}dx,
\end{equation*}
where for simplicity we have defined $G_p(x)=1/{\cal H}_p(x)$. We can derive identities for the $G_p(x)$ by taking the identities for ${\cal H}_p(x)$,
\begin{eqnarray*}
{\cal H}'_p(x)&={\cal H}_{p+1}(x)-2x{\cal H}_p(x)\cr
{\cal H}'_p(x)&=2p{\cal H}_{p-1}(x),
\end{eqnarray*}
and substituting ${\cal H}_p(x)=1/G_p(x)$ to obtain
\begin{eqnarray*}
G_p(x)^2&=2xG_p(x)G_{p+1}(x)-G'_p(x)G_{p+1}(x)\cr
G'_p(x)G_{p-1}(x)&=-2pG_p(x)^2.
\end{eqnarray*}
We can now substitute the first identity into the formula for $J_p(x)$ and integrate by parts to obtain
\begin{eqnarray*}
\fl
J_p(x)&=\int \left([2x e^{-x^2}]G_p(x)G_{p+1}(x)
-G'_p(x)G_{p+1}(x)e^{-x^2}\right)dx\cr
\fl
&=-G_p(x)G_{p+1}(x)e^{-x^2}+
\int \left([G_p(x)G_{p+1}(x)]'e^{-x^2}
-G'_p(x)G_{p+1}(x)e^{-x^2}\right)dx\cr
\fl
&=-G_p(x)G_{p+1}(x)e^{-x^2}+
\int G_p(x)G'_{p+1}(x)e^{-x^2}dx\cr
\fl
&=-G_p(x)G_{p+1}(x)e^{-x^2}-
2(p+1)\int G_{p+1}(x)^2 e^{-x^2}dx\cr
\fl
&=-G_p(x)G_{p+1}(x)e^{-x^2}
-2(p+1)J_{p+1}(x),
\end{eqnarray*}
where in the fourth step we used the second identity with $p$ replaced by $p+1$. Repeating the process and using the formula to evaluate the definite integral gives
\begin{equation*}
\fl
 J_p(x)\Big|_0^\infty=
[-G_p(x)+2(p+1)G_{p+2}(x)]G_{p+1}(x)e^{-x^2}\Big|_0^\infty
+4(p+1)(p+2)J_{p+2}(x)\Big|_0^\infty.
\end{equation*}
Suppose now that $p=2m$ is even.
The first term on the right hand side clearly vanishes as $x\rightarrow\infty$, but as $x\rightarrow 0$ the situation is more complicated since $G_{2m+1}(x)$ has an $O(1/x)$ singularity at $x=0$. From the identities for ${\cal H}_p(x)$ we see that
${\cal H}_{2m}(0)=2(2m-1){\cal H}_{2m-2}(0)$ from which we may deduce that ${\cal H}_{2m}(0)=(2m)!/m!$ and hence $G_{2m}(0)=m!/(2m)!$. The coefficient multiplying $G_{2m+1}(x)$ as $x\rightarrow 0$ is thus
\begin{equation*}
-{m!\over (2m)!}+2(2m+1){(m+1)!\over (2m+2)!}=0.
\end{equation*}
The $O(1/x)$ term thus has coefficient zero, and since the next term is $O(x)$, we see that the first term on the right hand side also vanishes as $x\rightarrow 0$. If we now multiply by $2$ to make the region of integration from $-\infty$ to $\infty$, we get the result
\begin{equation*}
I_{2m}=4(2m+1)(2m+2)I_{2m+2}.
\end{equation*}
Since $I_0=\sqrt{\pi}$, the result \eref{IntIdent} follows.

\section{Proof of two Hermite polynomial identities}
In this appendix we prove two identities involving Hermite and pseudo-Hermite polynomials, which are need to give an explicit form for eigenfunctions of generalized Cari\~nena potentials.

The first identity we prove is
\begin{equation}
\label{Ident1}
{\cal H}_p(x) H_k(x)=\sum_{i=0}^p
2^i {p!\over i!(p-i)!}{(k+p-i)!\over (k+p-2i)!}H_{k+p-2i}(x),
\end{equation}
of which the first few examples are
\begin{eqnarray}
\label{FirstExamples}
1\cdot H_k(x)&=H_k(x)\cr
2x\cdot H_k(x)&=H_{k+1}(x)+2kH_{k-1}(x)\cr
(4x^2+2)\cdot H_k(x)&=H_{k+2}+4(k+1)H_k(x)+4k(k-1)H_{k-2}(x).
\end{eqnarray}
To prove this identity we will need the identities
\begin{eqnarray}
\label{HermIdents}
2xH_k(x)&=H_{k+1}(x)+2kH_{k-1}(x)\cr
{\cal H}_{p+1}(x)&=2x{\cal H}_p(x)+2p{\cal H}_{p-1}(x),
\end{eqnarray}
which are easily derived directly from the relevant Rodrigues formulae. In fact the identity \eref{Ident1} was originally obtained by using the first of \eref{HermIdents} to derive the examples \eref{FirstExamples}, and thence spot the general pattern.

The proof is by induction on $p$. The case $p=0$ is trivially true; the case $p=1$ is simply the first of \eref{HermIdents}. It follows that the formula is true for $p=0$ and $p=1$. Now assume it is true for all values up to $p\ge 2$. The formula for $p+1$ can then be written
\begin{eqnarray}
\label{InductionEqn}
\fl
{\cal H}_{p+1}(x)H_k(x)&=
2x{\cal H}_p(x)H_k(x)+2p{\cal H}_{p-1}(x)H_k(x)\cr
&= {\cal H}_{p}(x)H_{k+1}(x)+2k{\cal H}_p(x)H_{k-1}(x)
+2p{\cal H}_{p-1}(x)H_k(x),
\end{eqnarray}
using the identities \eref{HermIdents}. The three terms in \eref{InductionEqn} take the form
\begin{eqnarray}
\label{IdentTerms1}
\fl
\quad\quad
{\cal H}_{p}(x)H_{k+1}(x)&=\sum_{i=0}^p
2^i {p!\over i!(p-i)!}{(k+p+1-i)!\over (k+p+1-2i)!}
H_{k+p+1-2i}(x)\cr
&=\sum_{i=0}^{p+1}
2^i  {p!\over i!(p+1-i)!}\{p+1-i\}
{(k+p+1-i)!\over (k+p+1-2i)!}
H_{k+p+1-2i}(x)\cr
\fl
2k{\cal H}_p(x)H_{k-1}(x)&=2k \sum_{i=0}^p
2^i {p!\over i!(p-i)!}{(k+p-1-i)!\over (k+p-1-2i)!}
H_{k+p-1-2i}(x)\cr
&=\sum_{i=0}^p
2^{i+1} {p!\over i!(p-i)!}\{k\}{(k+p-1-i)!\over (k+p-1-2i)!}
H_{k+p-1-2i}(x)\cr
\fl
2p{\cal H}_{p-1}(x)H_k(x)&=2p\sum_{i=0}^{p-1}
2^i {(p-1)!\over i!(p-1-i)!}{(k+p-1-i)!\over (k+p-1-2i)!}
H_{k+p-1-2i}(x)\cr
&=\sum_{i=0}^p 2^{i+1}{p!\over i!(p-i)!}\{p-i\}
{(k+p-1-i)!\over (k+p-1-2i)!}H_{k+p-1-2i}(x),
\end{eqnarray}
where we were able to extend the summation limit in the first and last terms because the additional terms are exactly zero in both cases. Combining the last two terms in \eref{IdentTerms1} then yields
\begin{eqnarray}
\label{IdentTerms2}
&\sum_{i=0}^p 2^{i+1}{p!\over i!(p-i)!}\{k+p-i\}
{(k+p-1-i)!\over (k+p-1-2i)!}H_{k+p-1-2i}(x)\cr
=&\sum_{i=0}^p 2^{i+1}{p!\over i!(p-i)!}
{(k+p-i)!\over (k+p-1-2i)!}H_{k+p-1-2i}(x)\cr
=&\sum_{i=1}^{p+1} 2^i {p!\over (i-1)!(p+1-i)!}
{(k+p+1-i)!\over (k+p+1-2i)!}H_{k+p+1-2i}(x)\cr
=&\sum_{i=0}^{p+1} 2^i  {p!\over i!(p+1-i)!}\{i\}
{(k+p+1-i)!\over (k+p+1-2i)!}H_{k+p+1-2i}(x)
\end{eqnarray}
where in the second step we shifted $i\rightarrow i-1$, and in the third step we were again able to extend the summation limit because the additional term was zero. Adding \eref{IdentTerms2} to the first term of \eref{IdentTerms1} gives
\begin{eqnarray*}
\fl
{\cal H}_{p+1}(x)H_k(x)&=\sum_{i=0}^{p+1}
2^i {p!\over i!(p+1-i)!}\{p+1\}
{(k+p+1-i)!\over (k+p+1-2i)!}H_{k+p+1-2i}(x)\cr
&=\sum_{i=0}^{p+1} 2^i {(p+1)!\over i!(p+1-i)!}
{(k+p+1-i)!\over (k+p+1-2i)!}H_{k+p+1-2i}(x),
\end{eqnarray*}
which is just \eref{Ident1} with $p$ replaced by $p+1$. This completes the proof by induction.

The second identity we prove is
\begin{equation}
\label{Ident2}
\fl
{\cal H}_p(x)H_{k+1}(x)+{\cal H}'_p(x)H_k(x)
=\sum_{i=0}^p 2^i {p!\over i!(p-i)!}\{k+p+1\}
{(k+p-i)!\over (k+p+1-2i)!}H_{k+p+1-i}(x)
\end{equation}
To prove this we note from \eref{Ident1} that
\begin{eqnarray}
\label{Ident2Term1}
\fl
{\cal H}_p(x)H_{k+1}(x)&=\sum_{i=0}^p
2^i {p!\over i!(p-i)!}{(k+p+1-i)!\over (k+p+1-2i)!}
H_{k+p+1-2i}(x)\cr
&=\sum_{i=0}^{p}
2^i  {p!\over i!(p-i)!}\{k+p+1-i\}{(k+p-i)!\over (k+p+1-2i)!}
H_{k+p+1-2i}(x),
\end{eqnarray}
whilst
\begin{eqnarray}
\label{Ident2Term2}
\fl
{\cal H}'_p(x)H_k(x)&=2p{\cal H}_{p-1}(x)H_k(x)\cr
&=2p\sum_{i=0}^{p-1} 2^i {(p-1)!\over i!(p-1-i)!}
{(k+p-1-i)!\over (k+p-1-2i)!}H_{k+p-1-2i}(x)\cr
&=\sum_{i=0}^{p-1} 2^{i+1} {p!\over i!(p-1-i)!}
{(k+p-1-i)!\over (k+p-1-2i)!}H_{k+p-1-2i}(x)\cr
&=\sum_{i=1}^p 2^i {p!\over (i-1)!(p-i)!}
{(k+p-i)!\over (k+p+1-2i)!}H_{k+p+1-2i}(x)\cr
&=\sum_{i=0}^p 2^i {p!\over i!(p-i)!} \{i\}
{(k+p-i)!\over (k+p+1-2i)!}H_{k+p+1-2i}(x),
\end{eqnarray}
where in the fourth step we shifted $i\rightarrow i-1$, and in the final step we were again able to extend the limit of summation. Adding \eref{Ident2Term1} and \eref{Ident2Term2} together then gives the required identity \eref{Ident2}.


\begin{thebibliography}{1}
\bibitem{Factor of 2}
Note that we define our Hamiltonian to be twice that used by Cari\~nena, so that our energy values are twice his. We use this definition so that the coefficient of the second derivative in Schr\"odinger's equation is -1, which is the standard choice in the factorization and supersymmetry literature.
\end{thebibliography}
\end{document}